\documentclass[9pt,twocolumn,twoside]{osajnl}
\journal{ol} % Choose journal (ao, aop, josaa, josab, ol, pr)

\setboolean{shortarticle}{true}
\title{Influence of optical feedback on harmonic pulsating solutions of long-cavity mode-locked VECSELs}

\author[1,$\dagger$]{A. Bartolo}
\author[2,3,$\dagger$]{T. Seidel}
\author[4]{N. Vigne}
\author[4]{A. Garnache}
\author[5]{G. Beaudoin}
\author[5]{I. Sagnes}
\author[1]{G. Huyet}
\author[1]{M. Giudici}
\author[2]{J. Javaloyes}
\author[2,3]{S. V. Gurevich}
\author[1,*]{M. Marconi}

\affil[1]{Université Côte d\textquoteright Azur, Centre National de La Recherche Scientifique, Institut de Physique de Nice, F-06560 Valbonne, France}
\affil[2]{Dpt. de F\'{\i}sica, Universitat de les Illes Balears \& Institute of Applied Computing and Community Code, Campus UIB, E-07122 Palma de Mallorca, Spain}
\affil[3]{Institute for Theoretical Physics \&  Center for Nonlinear Science (CeNoS), University of Münster, Schlossplatz 2, 48149 Münster, Germany}

\affil[4]{Institut d'Electronique et des Systèmes, UMR5214, Centre National de la Recherche Scientifique, University of Montpellier, 34000 Montpellier, France}
\affil[5]{Centre de Nanosciences et de Nanotechnologies, CNRS, Université Paris-Saclay, UMR 9001, 10 Boulevard Thomas Gobert, 91120, Palaiseau, France}

\affil[*]{Corresponding author: mathias.marconi@inphyni.cnrs.fr}
\affil[$\dagger$]{These two authors contributed equally}

\doi{\url{http://dx.doi.org/10.1364/XX.XX.XXXXXX}}
\ociscodes{140.4050,190.5530,190.4370}

\graphicspath{{./}{./plots/}} % to look in various

\begin{abstract}
	We analyse the effect of optical feedback on the dynamics of external-cavity mode-locked semiconductor lasers operated in the long cavity regime. Depending on the ratio between the cavity round-trip time and the feedback delay, we show experimentally that feedback acts as a solution discriminator that either reinforces or hinders the appearance of one of the multiple coexisting mode-locked harmonic solutions. Our theoretical analysis reproduces well the experiment. We identify asymmetrical resonance tongues due to the temporal symmetry breaking induced by gain depletion.

\end{abstract}
\setboolean{displaycopyright}{true}

\begin{document}
	
	\maketitle
	
%\section{Introduction}

Vertical External-Cavity Semiconductor lasers (VECSELs) allow to obtain stable, high output power lasing, with excellent beam qualities \cite{WTB-OE-13}. They allows for continuous wave (CW) speckle-free operation with a self-imaging cavity \cite{CCB-NAP-19}, wavelength tunability \cite{CZF-AO-18}, bi-frequency emission for THz \cite{BPM-JSTQE-17}, or pulsed operation when a saturable absorber (SA) is placed in the external cavity \cite{KT-PR-06,CSV-OL-18}.
In the latter case, VECSELs allow to obtain low repetition rate mode-locked lasers, with interesting applications in, e.g, dense frequency comb spectroscopy \cite{LMW-SCI-17,vangasse2020onchip}. Recent works have demonstrated that pulse trains with ultra-low period (hence spectrally dense comb) can be achieved from passively mode-locked (PML) VCSELs with SA in the so-called long cavity regime \cite{MJB-PRL-14}. In this situation, the pulses become independent, and addressable, temporal localized structures (TLSs).

The control, manipulation and optimization of the semiconductor PML lasers dynamics has become an extremely attractive topic due to its strong potential for applications. In particular, optical feedback improves the timing jitter in high repetition rates mode-locked lasers and offers the possibility to precisely harness the pulse train repetition rates \cite{OLV-NJP-12,AKB-APL-13,JPR-PRA-15,NJD-OE-16,JKL-CHA-17}. Recent works addressed the nonlocal interactions induced by a second delay on vectorial TLSs observed in VCSELs \cite{MJB-NPH-15,JMG-PRL-17}. However, the impact of feedback in PML semiconductor lasers operated in the long cavity regime still remains poorly understood, both experimentally and theoretically. To our knowledge, optical feedback was mainly studied in the context of spatial solitons were it leads to zigzagging \cite{PVG-PRA-16,STP-CHA-17}, drifts \cite{VPG-PTA-14}, pulsations \cite{SPT-PRA-17} or chaos \cite{PT-OL-14}.
\begin{figure}[htp]
	\includegraphics[scale=0.5,angle=0,origin=c]{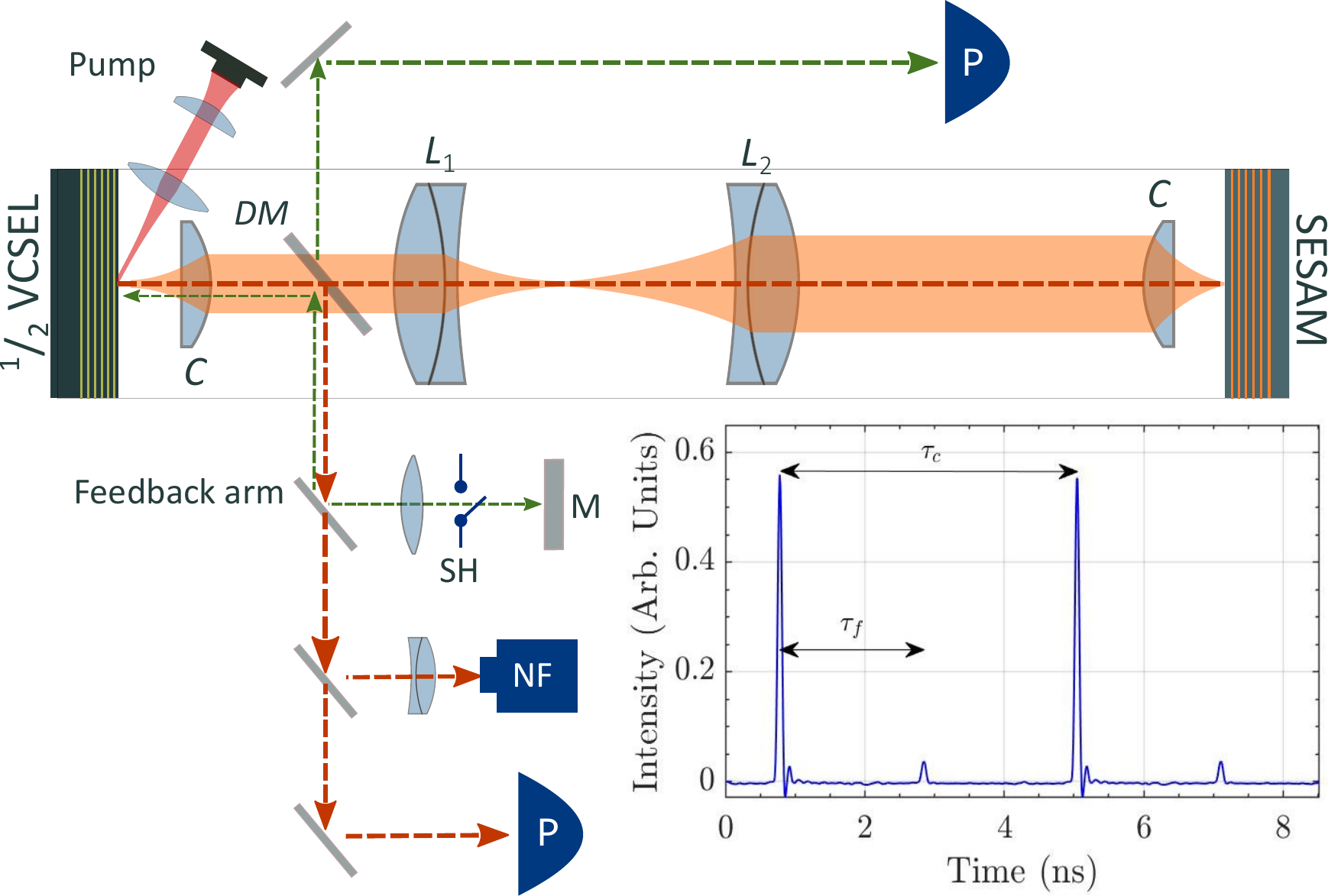}% Here is how to import EPS art
	\caption{Experimental setup of the PML VECSEL with optical feedback. DM: Dichroic Mirror, C: collimator, L1 and L2: achromatic lenses, P: Photodiode, NF: spatial near-field of the VECSEL emission recorded on a camera, M: Mirror, SH: mechanical shutter. Inset: Typical pulse train with period $\sim\tau_c$ and where the feedback induced satellite is located at $\Delta \tau=\tau_f-\tau_c$ from the main pulse. \label{fig:one}}
\end{figure}

In this paper, we address the effect of optical feedback in a low repetition rate PML VECSEL operating in the long cavity regime. Choosing a rational ratio between the feedback delay and the cavity round-trip allows selecting one of the multiple harmonic PML solutions that coexist close to the lasing threshold.
Due to the parity breaking effects \cite{JCM-PRL-16,PGG-NAC-20} in this system, we observe that the resonance tongues induced by the optical feedback are strongly asymmetrical, which has a clear interpretation in terms of an additional gain depletion occurring before or after the emission of a pulse. Our theoretical analysis is in excellent agreement with the experiment.    
%\section{Experimental results}

The experimental cavity configuration is shown in Fig.~\ref{fig:one}. The gain medium consists in 6 quantum wells embedded between a bottom totally reflective Bragg mirror and a top partially reflective Bragg mirror (\textonehalf{} VCSEL). We place the \textonehalf{} VCSEL in an external cavity that is closed by a fast  semiconductor saturable absorber mirror (SESAM) to operate the laser in the PML regime. The cavity is made long enough so that the laser operates in the TLS regime \cite{MJB-PRL-14,CSV-OL-18}. In this regime, the TLSs can be independently addressed by an external perturbation
\cite{CJM-PRA-16}. To avoid any spatial dynamics \cite{CCB-NAP-19}, the lenses are positioned in the cavity in order to insure that the output beam has a stable Gaussian profile. The regime of TLSs is characterized by a multistability close to the lasing threshold between the multiple harmonic mode-locking solutions (HML$_n$) in which the laser emits $n$ pulses separated by $\tau_c/n$.
The maximum number of pulses per round-trip $N_m$ is approximately equal to the ratio of the cavity round-trip $\tau_c$ and the gain recovery time $\tau_g$, $N_m=\tau_c/\tau_g$. For a cavity length of 63~cm $\tau_c = 4.26\,$ns and, since $\tau_g = 1\,$ns we find $N_m\simeq4$.

\begin{figure}
	\includegraphics[width=1\linewidth,height=0.7\linewidth,angle=0,origin=c]{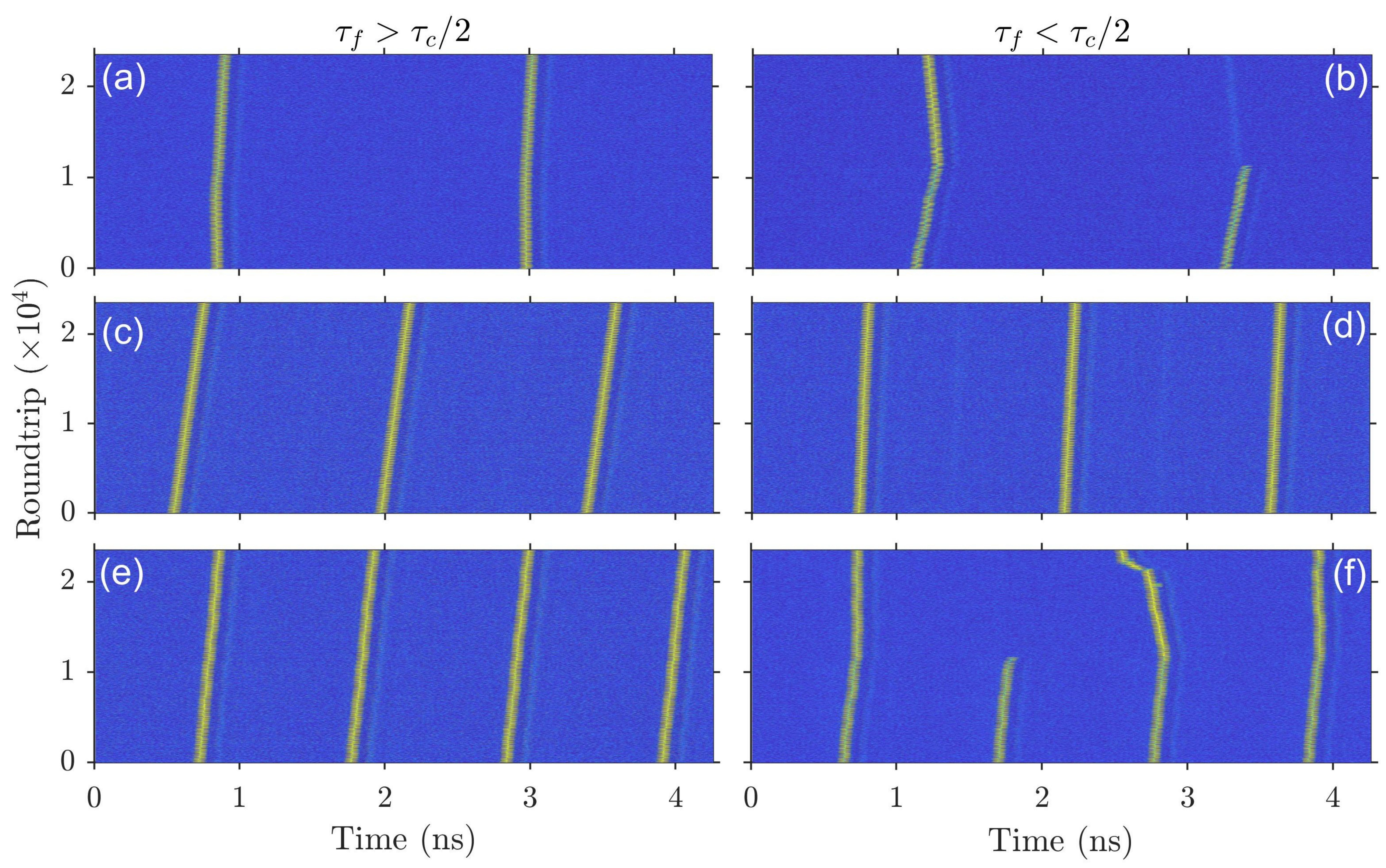}% Here is how to import EPS art
	\caption{Spatio-temporal diagrams of the pulse dynamics for two (top), three (middle) and four (bottom) TLSs in the cavity for two distinct values of the feedback delay $\tau_f$ around $\tau_c/2$.  \label{fig:two}} %a),b),c) Initial conditions with two, three and four TLSs, respectively. 
\end{figure}

To study the sensitivity of this regime to optical feedback, we implement a light re-injection arm closed by a mirror with 99 \% reflectivity at 1064~nm. To avoid diffraction losses in the feedback arm, we place a 50~mm lens that focuses the light on the mirror.  We use a mechanical shutter to open the feedback arm. The latter has a timescale of $\sim 100\,\mu$s which allows to conveniently ramp-up the feedback level.
With the presented setup one can reach a maximum of $0.03\%$ of feedback. This value is strongly limited by the intracavity dichroic mirror, that outcouples only $2$ \% of the light at 1064 nm. The remaining re-injected light that is transmitted by the dichroic mirror is collected by a fast photodiode and will be used to trigger the detection on the feedback signal when the arm is open. 

In the inset of Fig.~\ref{fig:one}, we show the effect of the feedback when it is applied in the regime of fundamental PML (one pulse per round-trip). The re-injected pulse interacts with the gain medium after a delay $\tau_f$ that is in this case smaller than $\tau_c$. One can see that the pulse will deplete the gain available during its interaction and produce a small pulse satellite at the distance $\Delta\tau=\tau_f-\tau_c$ for the main pulse; the latter is visible on the temporal time trace in  Fig.~\ref{fig:one}. Optical feedback is therefore inducing \emph{a nonlocal effect} via light-matter interaction in the gain medium. In the example of Fig.~\ref{fig:one}, the feedback does not perturb the fundamental PML regime since the satellite is re-injected far from an existing pulse, i.e. $\tau_f \neq n \tau_c$ and the feedback level is low.

We analyzed experimentally the effect of feedback when $\tau_f$ is slightly smaller or larger than the temporal interval between two consecutive pulses. We first present the results when $\tau_f \sim \tau_c/2$ and prepare the PML laser in the HML$_2$ regime. Depending on the precise value of $\tau_f$, the light pulse will be re-injected before or just after the emitted pulses.  
\begin{figure}
	\includegraphics[width=1\linewidth,height=0.7\linewidth,angle=0,origin=c]{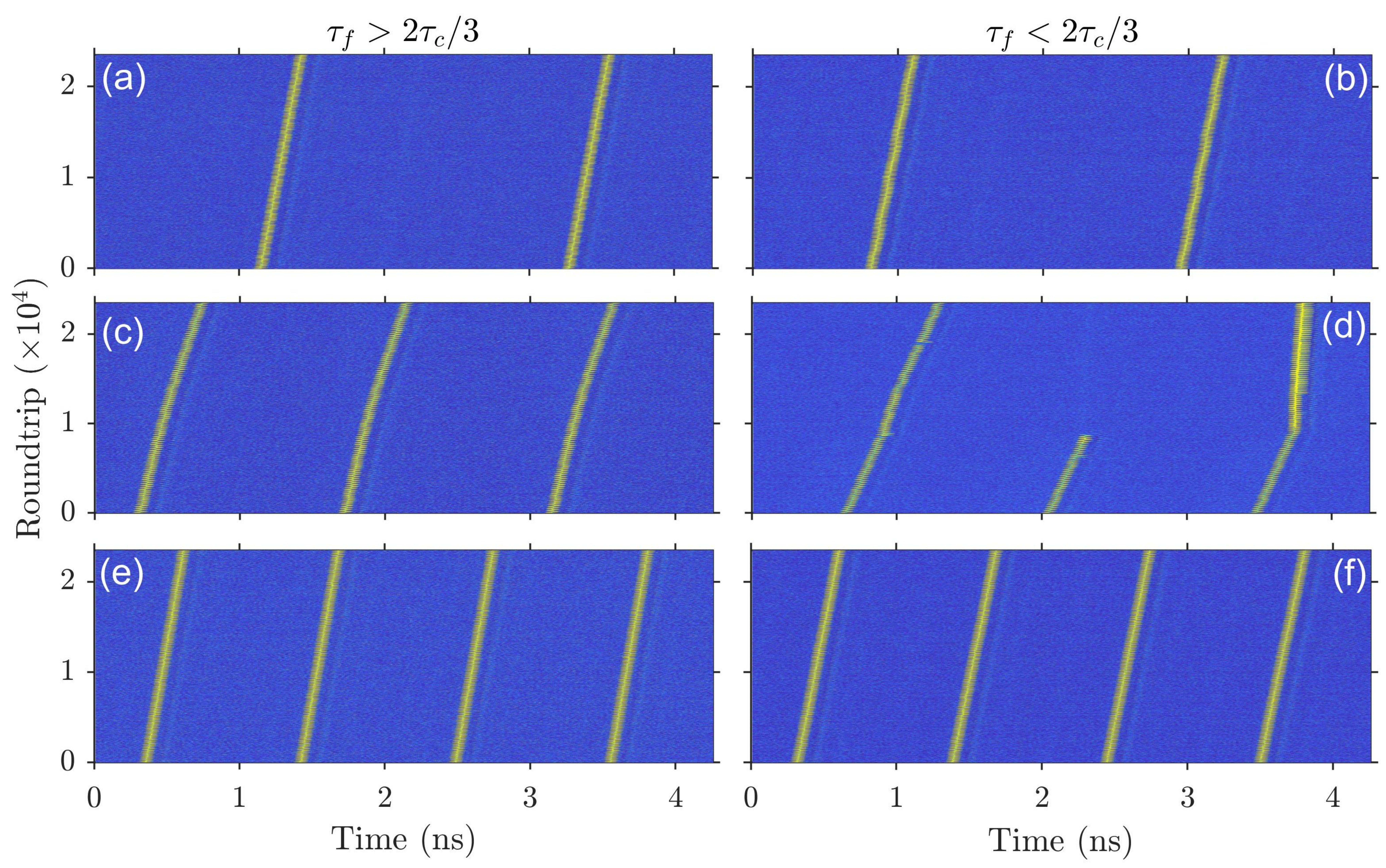}% Here is how to import EPS art
	\caption{Spatio-temporal diagrams of the pulse dynamics for two (top), three (middle) and four (bottom) TLSs in the cavity for two distinct values of the feedback delay around $2\tau_c/3$.  \label{fig:three}}%a) initial condition with 2 TLS. b) 3 TLS , c) 4 TLS.
\end{figure}
In Fig.~\ref{fig:two}a) we show the case where $\tau_f\gtrsim\tau_c/2$, with $\tau_f$ = 2.24~ns. The figure consists in a space-time representation \cite{GP-PRL-96} of the pulse propagation in the external cavity. There, consecutive chunks of duration $T$ of the time trace are stacked up vertically and $T\sim \tau_c$ corresponds to the period of the pulse train in the absence of feedback. The feedback starts to be applied at round-trip 0, where the laser is operating in a HML$_2$ regime with two equidistant pulses in the cavity separated by $\tau_c/2$. We observe that in this case, the feedback has no effect on the pulse dynamics. The situation changes drastically when we set $\tau_f$ = 2.06 ns, thus slightly smaller than $\tau_c/2$ (see Fig.~\ref{fig:two} b)). In this case we see that the feedback becomes eventually sufficient to erase one of the two pulses, leaving the system operating in the fundamental TLS regime. This effect can be understood quite straightforwardly. In fact, the re-injected feedback pulse is inducing an additional gain depletion. If it occurs just before another pulse reaches the gain section, it will lower the amplification of the latter. After several round-trips of diminished amplification, the pulse may eventually get erased.
After erasure, we also note that the remaining pulse is slowing down and its amplitude increases when the second one disappears, this is due to the increase of local gain experienced by the pulse in the cavity \cite{CJM-PRA-16}.
We now analyze the effect of the feedback when the laser is operating in the HML$_3$ and HML$_4$ regimes. Again, the feedback is ramped-up from  the first round-trip and $\tau_f$ is slightly larger (left column) or smaller (right column) than  $\tau_c/2$. Figures~\ref{fig:two}c,d) show that the feedback has no effect over the HML$_3$ regime independently of the precise value of $\tau_f$ chosen around $\tau_c/2$. Indeed, in this case the feedback-induced gain depletion occurs \emph{exactly} between two pulses that are already present in the cavity; the feedback depleted gain has enough time to relaxes to equilibrium and the other pulses do not feel this parasitic depletion. However, we observe that feedback can erase a pulse when starting from the  HML$_4$ solution, cf. Fig.~\ref{fig:two}e,f), in a way similar to the HML$_2$ case. This is easily explained by the fact that the HML$_4$ and HML$_2$ solutions both contain a pulse separated by $\tau_c/2$ on which the feedback with a delay slightly smaller than $\tau_c/2$ is going to act. 
After one pulse is erased, we clearly observe how the three remaining pulses start the process of rearrangement. This process is not captured until the end, but the space-time map gives an indication of the timescales at play. Finally, we conclude our experimental analysis by setting $\tau_f$ to another value this time around $2\tau_c/3$. Our results are shown in Fig.~\ref{fig:three} for the HML$_{2,3,4}$ solutions and $\tau_f$ slightly larger (left) or smaller (right) than $2\tau_c/3$. In this condition, we see that, as opposed to the previous case, the only solution that is being affected when $\tau_f \lesssim 2\tau_c/3$ is the HML$_3$ solution (Fig.~\ref{fig:three}d) while the even solutions HML$_{2,4}$ are not affected at all, cf. panels a),b),e),f). In Fig.~\ref{fig:three}d), we observe the same reconfiguration of the pulse positions in the cavity as in the $\tau_f<\tau_2/2$ case.
%\section{Theoretical Analysis}

To understand the experimental findings in details, we employ a widely used theoretical framework that considers a PML laser in a ring geometry in which the gain medium is coupled to a SA and a narrow band optical filter. Such a description is embodied in the delayed differential equation (DDE) model first presented in \cite{VT-PRA-05}. This model is extended by a term describing the time-delayed feedback as in \cite{OLV-NJP-12,JPR-PRA-15}. Denoting by $A$ the amplitude of the optical field, $G$ the gain, and $Q$ the saturable losses, the DDE model reads 
\begin{align}
\frac{\dot{A}}{\gamma} =&  \sqrt{\kappa}\exp\left[\frac{1-i\alpha_g}{2}G\left(t-\tau_{c}\right)-\frac{1-i\alpha_a}{2}Q\left(t-\tau_{c}\right)\right] \times \nonumber \\
&A\left(t-\tau_{c}\right)-A\left(t\right)+\eta e^{i\Omega}A\left(t-\tau_{f}\right),\label{eq:DDE_A}\\
%\frac{\dot{A}}{\gamma} =&  \sqrt{\kappa}\exp\left[\frac{1-i\alpha_g}{2}G_{\tau_c}-\frac{1-i\alpha_a}{2}Q_{\tau_{c}}\right] A{\tau_{c}}-A+\eta e^{i\Omega}A_{\tau_{f}},\label{eq:DDE_A}\\
\dot{G}  =& g_0-\Gamma G-e^{-Q}\left(e^{G}-1\right)\left|A\right|^{2},\label{eq:DDE_G}\\
\dot{Q} =&  q_{0}-Q-s\left(1-e^{-Q}\right)\left|A\right|^{2},\label{eq:DDE_Q}
\end{align}
where time has been normalized to the SA recovery time, $\tau_c$ is the cavity round-trip time, $\alpha_{g,a}$  are the linewidth enhancement factors of the gain and absorber sections, respectively, $\kappa$ the fraction of the power remaining in the cavity after each round-trip, $g_{0}$ the pumping rate, $\Gamma$ the gain recovery rate, $q_{0}$ is the value of the unsaturated losses which determines the modulation depth of the SA, $s$ the ratio of the saturation energy of the SA and of the gain sections and $\gamma$ is the bandwidth of the spectral filter, $\eta$ is the feedback rate, $\Omega$ is the feedback phase and $\tau_f$ the round-trip time of the feedback loop. The lasing threshold for resonant feedback reads 
$
g_{\text{th}}= \Gamma \left[q_0-\ln(\kappa)+2\cdot \ln(1-|\eta|)\right], \label{eq:g0th}
$
and we defined a normalized gain value $g=g_0/g_{\text{th}}$. We fix $(\gamma,\kappa,\Gamma,q_0,\alpha_g,\alpha_a,s,\eta,\Omega)=(10,0.8,0.04,0.3,1.5,0.5,10,0.005,0)$ while the cavity round-trip is set to $\tau_c=100$.

At $g=1$ the off state $(A,\,G,\,Q)=(0,\,g_0/\Gamma\,,q_0)$ becomes unstable and a branch of continuous wave (CW) solutions emerges. This branch undergoes several Andronov-Hopf (AH) bifurcations from which the fundamental (FML) and the HML$_{n}$ solutions emerge~\cite{MJB-PRL-14}. In the long delay limit the AH bifurcations become subcritical and eventually the branches of pulsating solutions detach from the CW branch. In this case, the latter may extend below the lasing threshold and coexist with the off state~\cite{MJB-PRL-14}. There, the localized solutions gain stability via Saddle-Node bifurcation of Limit Cycles (SN) for the FML solution or a Torus bifurcation for HML$_n$ solutions.
\begin{figure}
	\centering\includegraphics[width=1\linewidth]{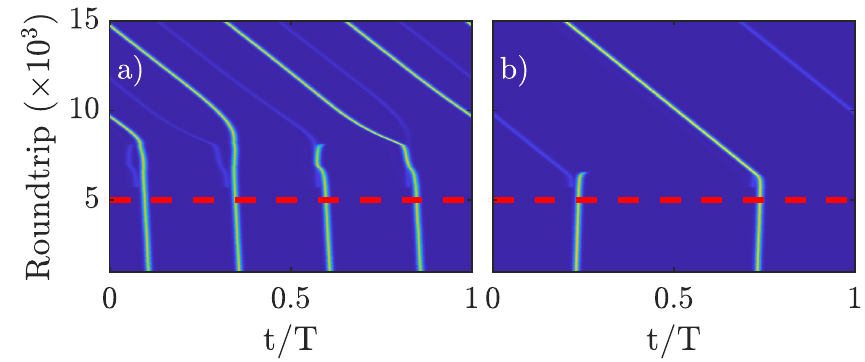}
	\caption{Direct numerical simulations of Eqs.~(\ref{eq:DDE_A}-\ref{eq:DDE_Q}) for the HML$_4$ a) and HML$_2$ b) solutions. Time delayed feedback is turned on after 5000 round-trips as indicated by the dashed red line. The gain value are a) $g=1.08$, b) $g=0.952$ and $\tau_f=48$. }
	\label{fig:space_time}
\end{figure}
The direct numerical simulations of Eqs.~(\ref{eq:DDE_A}-\ref{eq:DDE_Q}) displayed in Fig.~\ref{fig:space_time} reproduce well the experimental findings depicted in Figs.~\ref{fig:two},\ref{fig:three}.
% for $\tau_f$ in the same regime as in the experiment.
% in which we matched the value of the time delays ratio $\tau_f/\tau_c$.
The system is initialized with the HML$_4$ and the HML$_2$ solutions and optical feedback is applied after 5000 roundtrips (red dashed line); in both cases feedback destroys the HML$_n$ solution and the system settles after a transient on a HML$_{n-1}$ solutions instead. This begs the question in which regimes time-delayed feedback has a destabilizing effect on the TLSs. A detailed bifurcation analysis using path continuation techniques was performed employing the software package \texttt{DDE-BIFTOOL}~\cite{DDEBT} 
which can follow solutions in parameter space, continue bifurcation points in two-parameter planes and allows to determine the stability of periodic solutions by computing their Floquet multipliers $\mu$.
The normalized gain $g$ is used as the main continuation parameter while the solution measure is $P=\max(|A|^2)$. Figures \ref{fig:bif_diagram}a)-c) show bifurcation diagrams in $g$ for a HML$_4$ solution for different values of $\tau_f$ close to $\tau_c/2$. In Fig.~\ref{fig:bif_diagram}a) the satellite is placed close to the leading edge of the main pulse as in Fig.~\ref{fig:space_time}a). As expected the branch is unstable for a wide range of $g$ because the satellite depletes the gain which cannot recover fast enough before the main pulse arrives. Only for high gain values when enough amplification is provided for both the satellite and the main pulse, the solution restabilizes via a torus bifurcation $H$ (green square). 
When the time-delayed feedback is applied resonantly as seen in Fig.~\ref{fig:bif_diagram}~b), i.e. when the satellite coincides with the main pulse, the range of stability increases significantly. Placing the satellite at the trailing edge of the main pulse as in Fig.~\ref{fig:bif_diagram}c) does not destabilize the solution. However, the range of stability is slightly smaller than in the resonant case. 

To quantify these results even further, it is helpful to consider the $(\tau_f,g)$-plane which is displayed in Fig.~\ref{fig:bif_diagram}d). Here, the colormap encodes the absolute value of the maximal Floquet multiplier for a given HML$_4$ solution. It is obtained by following 104 branches in $g$ of approximately 80 steps each for different (non-uniformly distributed) values of $\tau_f$. After computing the Floquet multipliers for each periodic solution, the data were interpolated. The white contour line in Fig.~\ref{fig:bif_diagram}d) represents the border of stability of the HML$_4$ solution.
\begin{figure}
	\centering\includegraphics[width=1\linewidth]{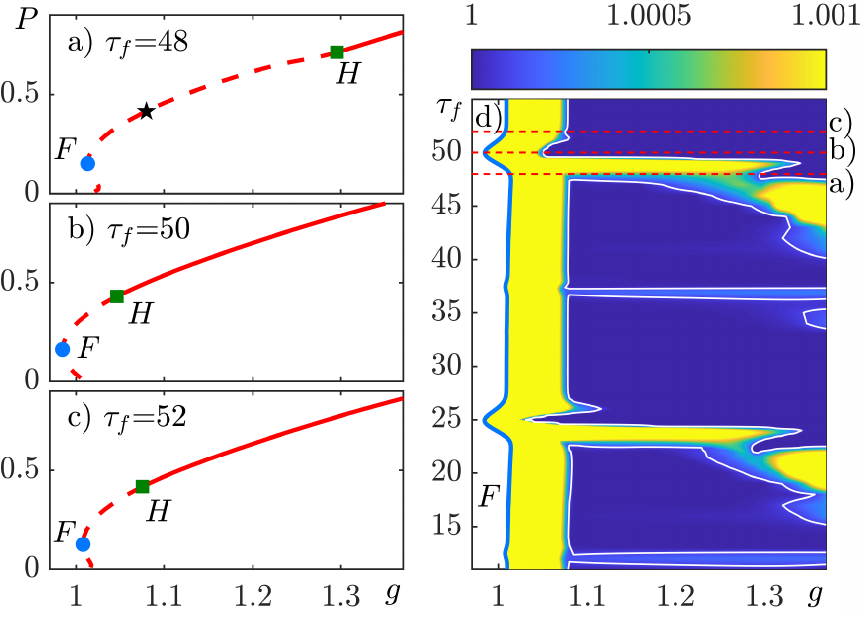}
	\caption{Bifurcation diagrams of eqs. (\ref{eq:DDE_A},\ref{eq:DDE_Q}) for a HML$_4$ solution. a)-c) Solid (resp. dashed) lines stand for stable (resp. unstable) solutions. The black star in a) indicates the parameters in Fig.~\ref{fig:space_time}a), the blue dots mark SN bifurcations and the green squares denote the Torus bifurcations in which the solutions gain stability. d) Analysis in the $(\tau_f,g)$-plane. The blue line on the left marked with $F$ corresponds to the SN bifurcation marked in a)-c). The dashed red lines indicate the cross-sections at which a)-c) are plotted. The colormap shows $\mathrm{max}(|\mu|)$. Blue regions within the white contour line correspond to stable regions while the rest is unstable.}
	\label{fig:bif_diagram}
\end{figure}
It can be clearly seen that the region in which time-delayed feedback destabilizes the HML solution is {\it asymmetrical} and limited to the vicinity of the leading edge of the main pulse ($47.5\lesssim\tau_f\lesssim49.4$). When the satellite is placed even closer to the main pulse, the opposite is the case and the satellite increases the range of stable HML solutions, which can be seen in form of a bump in the colormap at $\tau_f=\tau_c/2=50$. The bump also occurs in the line corresponding to the SN bifurcation $F$. Placing the satellite on the trailing edge of the main pulse does not influence the dynamics as the satellite is unable to interact with the main pulse. As the displayed branches correspond to a HML$_4$ solution one can expect to see more resonances in the ($\tau_f,g$)-plane at $\frac{\tau_f}{\tau_c}=\frac{1}{4},\frac{3}{4}, 1,\text{etc}$. Indeed, at $\tau_f=25$ a similar resonance in form of a bump in the colormap and the SN line can be observed. Interestingly, it has a different shape compared to the $\tau_f=50$, since $\tau_f=25$ couples all main pulses to their neighboring main pulse while $\tau_f=50$ creates two interspersed pairs of coupled pulses. Another peculiar property can be seen at $\tau_f=12.5$ and $\tau_f=37.5$. At these points second order resonance can be observed, meaning that the satellite's satellite destabilizes the main pulse for a short range. The SN line also exhibits small bumps for these values.\\
In conclusion, our analysis shows that coherent optical feedback acts as a reliable discriminator between the various multistable HML$_n$ solutions that coexist in a long-cavity PML VECSEL. Provided that feedback is applied slightly before one of the pulse present in the cavity, it may hinders the appearance of the associated HML$_n$ solution. Our results are well reproduced by a DDE model for PML including delayed feedback. A two-parameter bifurcation analysis exhibits strongly asymmetrical resonances around $\tau_f=\tau_c/n$ that are the result of the breaking of the temporal inversion symmetry due to gain depletion.
Further works will investigate in detail the effect of feedback in the resonance tongues disclosed in this work.
\section*{funding}
Ministerio de Economía y Competitividad (MOVELIGHT PGC2018-099637-B-100);Deutsche Forschungsgemeinschaft (B5 of the SFB 787);Deutscher Akademischer Austauschdienst PRIME programme;
French RENATECH network; ANR-18-CE24-0002 BLASON and Conseil R\'egional Provence-Alpes-C\^ote d’Azur (plateforme OPTIMAL).
\section*{Disclosures}
The authors declare no conflicts of interest.
	
%\bibliography{full,extra}

\begin{thebibliography}{10}
\newcommand{\enquote}[1]{``#1''}

\bibitem{WTB-OE-13}
K.~G. Wilcox, A.~C. Tropper, H.~E. Beere, D.~A. Ritchie, B.~Kunert, B.~Heinen,
  and W.~Stolz, \enquote{4.35 kw peak power femtosecond pulse mode-locked
  vecsel for supercontinuum generation,} Opt. Express \textbf{21}, 1599--1605
  (2013).

\bibitem{CCB-NAP-19}
H.~Cao, R.~Chriki, S.~Bittner, A.~A. Friesem, and N.~Davidson, \enquote{Complex
  lasers with controllable coherence,} Nature Reviews Physics \textbf{1},
  156--168 (2019).

\bibitem{CZF-AO-18}
B.~Chomet, J.~Zhao, L.~Ferrieres, M.~Myara, G.~Guiraud, G.~Beaudoin, V.~Lecocq,
  I.~Sagnes, N.~Traynor, G.~Santarelli, S.~Denet, and A.~Garnache,
  \enquote{High-power tunable low-noise coherent source at 1.06$\mu$m based on
  a surface-emitting semiconductor laser,} Appl. Opt. \textbf{57}, 5224--5229
  (2018).

\bibitem{BPM-JSTQE-17}
S.~Blin, R.~Paquet, M.~Myara, B.~Chomet, L.~L. Gratiet, M.~Sellahi,
  G.~Beaudoin, I.~Sagnes, G.~Ducournau, P.~Latzel, J.~F. Lampin, and
  A.~Garnache, \enquote{Coherent and tunable thz emission driven by an
  integrated iii-v semiconductor laser,} IEEE Journal of Selected Topics in
  Quantum Electronics \textbf{23}, 1--11 (2017).

\bibitem{KT-PR-06}
U.~Keller and A.~C. Tropper, \enquote{Passively modelocked surface-emitting
  semiconductor lasers,} Physics Reports \textbf{429}, 67 -- 120 (2006).

\bibitem{CSV-OL-18}
P.~Camelin, C.~Schelte, A.~Verschelde, A.~Garnache, G.~Beaudoin, I.~Sagnes,
  G.~Huyet, J.~Javaloyes, S.~V. Gurevich, and M.~Giudici, \enquote{Temporal
  localized structures in mode-locked vertical external-cavity surface-emitting
  lasers,} Opt. Lett. \textbf{43}, 5367--5370 (2018).

\bibitem{LMW-SCI-17}
S.~M. Link, D.~J. H.~C. Maas, D.~Waldburger, and U.~Keller, \enquote{Dual-comb
  spectroscopy of water vapor with a free-running semiconductor disk laser,}
  Science  (2017).

\bibitem{vangasse2020onchip}
K.~V. Gasse, Z.~Chen, E.~Vicentini, J.~Huh, S.~Poelman, Z.~Wang, G.~Roelkens,
  T.~W. Hänsch, B.~Kuyken, and N.~Picqué, \enquote{An on-chip
  iii-v-semiconductor-on-silicon laser frequency comb for gas-phase molecular
  spectroscopy in real-time,}  (2020).

\bibitem{MJB-PRL-14}
M.~Marconi, J.~Javaloyes, S.~Balle, and M.~Giudici, \enquote{How lasing
  localized structures evolve out of passive mode locking,} Phys. Rev. Lett.
  \textbf{112}, 223901 (2014).

\bibitem{OLV-NJP-12}
C.~Otto, L.~K., A.~Vladimirov, M.~Wolfrum, and E.~Schöll, \enquote{Delay
  induced dynamics and jitter reduction of passively mode-locked semiconductor
  laser subject to optical feedback,} New J. Phys. \textbf{14}, 113033 (2012).

\bibitem{AKB-APL-13}
D.~Arsenijević, K.~M., and B.~D., \enquote{Phase noise and jitter reduction by
  optical feedback on passively mode-locked quantum-dot lasers,} Appl. Phys.
  Lett. \textbf{103}, 231101 (2013).

\bibitem{JPR-PRA-15}
L.~Jaurigue, A.~Pimenov, D.~Rachinskii, E.~Sch\"oll, K.~L\"udge, and A.~G.
  Vladimirov, \enquote{Timing jitter of passively-mode-locked semiconductor
  lasers subject to optical feedback: A semi-analytic approach,} Phys. Rev. A
  \textbf{92}, 053807 (2015).

\bibitem{NJD-OE-16}
N.~O., L.~C. Jaurigue, L.~Drzewietzki, K.~Lüdge, and S.~Breuer,
  \enquote{Experimental demonstration of change of dynamical properties of a
  passively mode-locked semiconductor laser subject to dual optical feedback by
  dual full delay-range tuning,} Opt. Express \textbf{24}, 14301 (2016).

\bibitem{JKL-CHA-17}
L.~Jaurigue, B.~Krauskopf, and K.~Lüdge, \enquote{Multipulse dynamics of a
  passively mode-locked semiconductor laser with delayed optical feedback,}
  Chaos: An Interdisciplinary Journal of Nonlinear Science \textbf{27}, 114301
  (2017).

\bibitem{MJB-NPH-15}
M.~Marconi, J.~Javaloyes, S.~Barland, S.~Balle, and M.~Giudici,
  \enquote{Vectorial dissipative solitons in vertical-cavity surface-emitting
  lasers with delays,} Nat. Photon. \textbf{9}, 450--455 (2015). Article.

\bibitem{JMG-PRL-17}
J.~Javaloyes, M.~Marconi, and M.~Giudici, \enquote{Nonlocality induces chains
  of nested dissipative solitons,} Phys. Rev. Lett. \textbf{119}, 033904
  (2017).

\bibitem{PVG-PRA-16}
D.~Puzyrev, A.~G. Vladimirov, S.~V. Gurevich, and S.~Yanchuk,
  \enquote{Modulational instability and zigzagging of dissipative solitons
  induced by delayed feedback,} Phys. Rev. A \textbf{93}, 041801 (2016).

\bibitem{STP-CHA-17}
T.~Schemmelmann, F.~Tabbert, A.~Pimenov, A.~G. Vladimirov, and S.~V. Gurevich,
  \enquote{Delayed feedback control of self-mobile cavity solitons in a
  wide-aperture laser with a saturable absorber,} Chaos: An Interdisciplinary
  Journal of Nonlinear Science \textbf{27}, 114304 (2017).

\bibitem{VPG-PTA-14}
A.~G. Vladimirov, A.~Pimenov, S.~V. Gurevich, K.~Panajotov, E.~Averlant, and
  M.~Tlidi, \enquote{Cavity solitons in vertical-cavity surface-emitting
  lasers,} Phil. Trans. R. Soc. A \textbf{372}, 20140013 (2014).

\bibitem{SPT-PRA-17}
C.~Schelte, K.~Panajotov, M.~Tlidi, and S.~V. Gurevich, \enquote{Bifurcation
  structure of cavity soliton dynamics in a vertical-cavity surface-emitting
  laser with a saturable absorber and time-delayed feedback,} Phys. Rev. A
  \textbf{96}, 023807 (2017).

\bibitem{PT-OL-14}
K.~Panajotov and M.~Tlidi, \enquote{Chaotic behavior of cavity solitons induced
  by time delay feedback,} Opt. Lett. \textbf{39}, 4739--4742 (2014).

\bibitem{JCM-PRL-16}
J.~Javaloyes, P.~Camelin, M.~Marconi, and M.~Giudici, \enquote{Dynamics of
  localized structures in systems with broken parity symmetry,} Phys. Rev.
  Lett. \textbf{116}, 133901 (2016).

\bibitem{PGG-NAC-20}
A.~M. Perego, B.~Garbin, F.~Gustave, S.~Barland, F.~Prati, and G.~J.
  de~Valc\'{a}rcel, \enquote{Coherent master equation for laser modelocking,}
  Nat. Commun. \textbf{11}, 311 (2020).

\bibitem{CJM-PRA-16}
P.~Camelin, J.~Javaloyes, M.~Marconi, and M.~Giudici, \enquote{Electrical
  addressing and temporal tweezing of localized pulses in passively-mode-locked
  semiconductor lasers,} Phys. Rev. A \textbf{94}, 063854 (2016).

\bibitem{GP-PRL-96}
G.~Giacomelli and A.~Politi, \enquote{Relationship between delayed and
  spatially extended dynamical systems,} Phys. Rev. Lett. \textbf{76},
  2686--2689 (1996).

\bibitem{VT-PRA-05}
A.~G. Vladimirov and D.~Turaev, \enquote{Model for passive mode locking in
  semiconductor lasers,} Phys. Rev. A \textbf{72}, 033808 (2005).

\bibitem{DDEBT}
K.~Engelborghs, T.~Luzyanina, and D.~Roose, \enquote{Numerical bifurcation
  analysis of delay differential equations using dde-biftool,} ACM Trans. Math.
  Softw. \textbf{28}, 1--21 (2002).

\end{thebibliography}
%\bibliographyfullrefs{full,extra}

\end{document}